\documentclass[twocolumn,amsmath,amssymb,floatfix,
			   superscriptaddress]{revtex4}
\usepackage{url}
\usepackage{subfig}
\usepackage{graphicx}

\begin{document}
	
	\title{Network Structure Revealed by Short Cycles}
	\author{James~Bagrow}
		\email{bagrowjp@clarkson.edu} 
		\affiliation{Department of Physics, Clarkson 
		University, Potsdam, NY 13699-5820, USA.} 

	\author{Erik~Bollt}
		\email{bolltem@clarkson.edu} 
		\affiliation{Department of Mathematics and Computer
		Science, Clarkson University, Potsdam, NY 13699-5815,
		USA. }
		\affiliation{Department of Physics, Clarkson
		University, Potsdam, NY 13699-5820, USA.}

	\author{Luciano da F. Costa}
		\email{luciano@if.sc.usp.br} 
		\affiliation{Instituto de F\'{\i}sica de  S\~ao
		Carlos. Universidade de S\~ao Paulo, S\~ao Carlos, SP,
		PO Box 369, 13560-970, Brazil}

	\date{\today} 	
	%\keywords{Complex Networks; Communities; Cycles}
	%\pacs{}%23.23.+x, 56.65.Dy}

\begin{abstract} This article explores the relationship between 
communities and short cycles in complex networks, based on the 
fact that nodes more densely connected amongst one another are more 
likely to be linked through short cycles. By identifying combinations 
of 3-, 4- and 5-edge-cycles, a subnetwork is obtained which contains 
only those nodes and links belonging to such cycles, which can then 
be used to highlight community structure. Examples are shown using a 
theoretical model (Sznajd networks) and a real-world network 
(NCAA football). 
\end{abstract}
	
\maketitle

\section{Introduction} 
Complex networks have attracted growing attention because of their
non-uniform connectivity patterns, which may give rise to node degree
power laws and hubs, known to play an important role in defining several 
topological properties of the networks~\cite{ModelingInternetTopology,
dorogovtsev-2002-51,HandbookGraphsNetworks}. More recently, the fact 
that many complex networks include \emph{communities}, i.e. sets of nodes
which connect more intensely amongst one another than with the rest of 
the network, has become the focus of increasing attention
(e.g.~\cite{NewmanGirvan:Communities,NewmanGirvan:MixingPatterns:book:2003,Newman:2004:EPJ,Newman:2004:PRE, Newman:FastCommunities1:PRE,NewmanClausetMoor:FastModularityPRE:2004,BagrowBollt:LCD,Newman:PNAS:review:2006, Newman:EigenvectorsCommunities:PRE:2006,PorterMuchaNewman:USCongress}). Indeed, because of
statistical fluctuations, even random
networks~\cite{Erdos:RandomGraphs,bollobas:randomGraphs} can be found
to exhibit communities ~\cite{Guimera:ModularityRandom,
Reichardt:arbitraryCommunities}.  Although we still lack a clear-cut
definition of a community, the problem of identifying communities in
complex networks continues to motivate interest from researchers
because of the importance that those structures have for better
understanding the general organization of such complex structures
(e.g.~\cite{Amaral_Nature:2005}).

Another important feature of complex networks are the cycles of
different lengths which underlie the connectivity of the several
models of networks~\cite{Bollt_Avraham:2005}. Actually, the
statistical distribution of cycles has been acknowledged as
particularly important for defining not only the topology of the
respective networks, but also the dynamics of systems running on such
frameworks(e.g.~\cite{Arenas:2006}). The latter is a direct
consequence of the fact that cycles, through feedback, form the
scaffolding of memory in dynamical systems.
	
Generally, the density of cycles tends to increase as more edges are
incorporated into a network, with longer cycles being observed earlier
than shorter ones (e.g.~\cite{Costa:2004}). Therefore, the density of
cycles of different lengths can be used as an indicator of the
connectivity between any subset of nodes. In other words, the larger
the number of shortest cycles among a subset of nodes, the more
connected such nodes are to one another.  Longer cycles tend to grow,
``coiled up'', alongside these shorter cycles, however, blurring the
distinction between nodes based solely on short-cycle participation.
We present methods to overcome this.
	
The article starts by presenting the cycle finding algorithm and its
application as the core of the community finding algorithm and
proceeds by illustrating the application of such a methodology to
community finding in a theoretical complex network model (i.e. Sznajd
networks~\cite{Costa_Sznajd:2005}) and a real-world football network.

\section{Describing Short Cycles} 

For a graph $G = \left\{ V, E \right\}$, $n=|V|, m=|E|$, we are
interested in finding cycles of length 3,4, or 5 containing some
starting vertex $v \in V$. To describe these cycles we begin by
decomposing G into shells $S_i$ about $v$. We define shell $S_i$ to be
the set of all vertices (and edges between those vertices) at a
distance $i$ from the starting vertex $v$. Since we are only
interested in cycles of length $\leq 5$, we need only to keep shells
$S_1$ and $S_2$.
	
It is simple to describe all possible short cycles using these shell 
decompositions. For example, for every edge $e_{ij}$ in $S_1$ about 
$v$, there exists a 3-cycle (triangle) $v$--$i$--$j$--$v$. Similarly, 
for every path of length 2 or 3 in $S_1$, there exists a 4- or 
5-cycle, respectively. Another 4-cycle and two more 5-cycles exist 
involving both $S_1$ and $S_2$. %See Figure \ref{shell_cycles.fig}.
	
In general, for a cycle of length $L\geq3$, the number of such
possible ``cases'' grows with $L$.  Since it requires 2 edges to visit
a shell, an $L$-cycle can visit at most $J$ shells, where
\begin{equation}
	J = \left\{ \begin{array}{ll}
	\frac{L}{2}, \quad  & L~\text{even},\\
	\newline \\
	\frac{L-1}{2}, \quad & L~\text{odd}.
	\end{array} \right.
\end{equation}	

If the farthest shell the cycle visits is $S_j$, $j<J$, there are at
most $L-2j$ remaining edges that must be distributed between and
within the $S_1,S_2,...S_j$ shells.  The number of ways to distribute
$L-2j$ edges over $j$ shells is $\frac{(L-2j+j-1)!}{(L-2j)!(j-1)!}$.
However, it is possible for a cycle to ``zig-zag'' between shells,
using more than the $2j$ edges necessary to visit the $j$ shells.
Therefore, the total number of possible ways to distribute an
$L$-cycle is at least:

\begin{eqnarray}
	\lefteqn{N_{l}(L) =  1+} \nonumber \\ 
	& & \sum_{j=2}^{J} \sum_{i=0}^{J-j} \frac{(i+j-2)!}{i!(j-2)!} \frac{(L-2i-j-1)!}{(L-2j-2i)!(j-1)!} ,
\end{eqnarray}
with the outer sum accounting for all the possible shells the cycle 
can visit, the inner sum for all the optional pairs of edges that can 
lie between shells and the $+1$ for the one possible cycle that visits 
the first shell only.  Here, $i$ is the number of pairs of edges 
between shells beyond the $j$ necessary to visit the $j$ shells.

This calculation fails to take into account permutations of the 
\emph{ordering} of edges between and within two adjacent shells.  
A simple upper bound is possible, however, as there are certainly 
no more than $L!$ possible permutations over the whole network:
\begin{eqnarray}
	\lefteqn{N_{u}(L) = 1+} \nonumber \\
	& & \sum_{j=2}^{J} \sum_{i=0}^{J-j} \frac{(i+j-2)!}{i!(j-2)!} \frac{(L-2i-j-1)!}{(L-2j-2i)!(j-1)!}L!,
\end{eqnarray}
with
\begin{equation}
	N_{l}(L) \leq N(L) \leq N_{u}(L).
\end{equation}

\section{Cycles and Communities}\label{sec:cycles_communities} 
For a graph $G$, a cycle $C$ is a subset of the set of edges $E$ 
containing a continuous path, where the first and last node of the 
path are the same~\cite{bollobas_MGT}.  Permutations of cycles may 
be ignored since we will be working exclusively with sets of edges.  
Throughout this work, we limit ourselves to short cycles, typically 
those of length $l$, $3\leq l <6$.  These shorter cycles may provide 
the advantage of faster calculation times.

Community structure can be studied by comparing the edges covered by 
these cycles with the original graph. Let 
\begin{eqnarray}
  C_l(i) &\equiv & \mbox{the set of edges traversed by all} \\
  \nonumber & & \mbox{$l$-cycles starting from vertex $i$} 
\end{eqnarray}

Starting from all vertices and limiting ourselves to only short 
$j$-cycles \footnote{\protect{Indeed, here we specify short cycles as 
those of length 3, 4, or 5 but this is not a set rule and, in certain 
circumstances, it may prove advantageous to consider 4- or 5-cycles, 
or even just 5-cycles.}}, 
\begin{eqnarray}
  C & \equiv & \bigcup_{i \in V} \bigcup_{j} C_j(i) .
\end{eqnarray}

Then, for a graph $G$, we construct a graph $H$ where, 
\begin{eqnarray}
  H & = & \left\{ V, E \setminus C \right\} 
\end{eqnarray}
is the graph $G$ containing only edges that do not participate 
in $j$-cycles. Separate communities in $G$ will appear as disconnected 
components in $H$. We interpret vertices with degree zero in $H$ as 
communities of size one.

In specifying $H$, the question of what to choose for $j$ has been left 
open.  For example, choosing just $j=\{3\}$ will correspond to deleting 
all edges from $G$ that participate in 3-cycles, generally not a useful 
result.  One may consider $j$ to be a tunable parameter, used to get a 
desired result when applied to a specific network.
	
One issue that can occur is that longer cycles often overlap shorter 
cycles. In terms of communities, most inter-community edges contain 
few (if any) short cycles, but intra-community edges tend to contain 
both long and short cycles, since a long cycle can ``coil'' 
inside the community. If one were to just delete all 5-cycles in a graph, 
it is very possible to end up deleting all edges.  
	
There is quite a bit of leeway in how we choose $j$ and build $H$, and 
we can use this to our advantage.  For example, pick two cycle lengths 
$s$ and $t, s<t$ and compute $C_s$ and $C_t$.  Then, build another set 
of edges, $C_{t \setminus s}$
\begin{equation}
  C_{t \setminus s} \equiv C_t \setminus C_s,
  \label{eqn_t_not_s}
\end{equation}
containing the set of edges that participate in $t$-cycles \emph{but not} 
$s$-cycles.  The graph $H = \{V, C_{t \setminus s}\}$ will contain edges 
that tend to be between communities and not within, for an appropriate 
choice of $t$ and $s$.  One can think of this as a ``backbone'' of the 
network, and deleting these edges may be a useful pre-processing step
for applying other community-detection algorithms, including 
betweenness~\cite{NewmanGirvan:Communities,BagrowBollt:LCD}.

\section{Application Examples}\label{sec:application examples}  
We present example applications of the methods presented in Section
\ref{sec:cycles_communities} to two networks: a network of NCAA 
Division I-A football games held during the 2005 regular season
\footnote{Data taken from published schedule at
\protect\url{http://www.ncaa.org}} and a Sznajd network~\cite{Sznajd:2000}.  
In addition, we discuss how these methods can break down and ways to
overcome that.

\subsection{Football Network}
In NCAA football, teams are grouped into \emph{conferences} based on
location.  To save on transportation time and cost, more games are
played between teams in the same conference than in different
conferences.  Thus, a graph of the game schedule, where nodes are
teams and edges connect teams that have played against each other,
naturally exhibits community structure based on these conferences
\cite{MuchaFootBallCommunities}.
	
\begin{figure*}
	\centering
	\subfloat[]{
	    \label{NCAA_2005_orig.fig}
	    \includegraphics[width=0.48\textwidth]{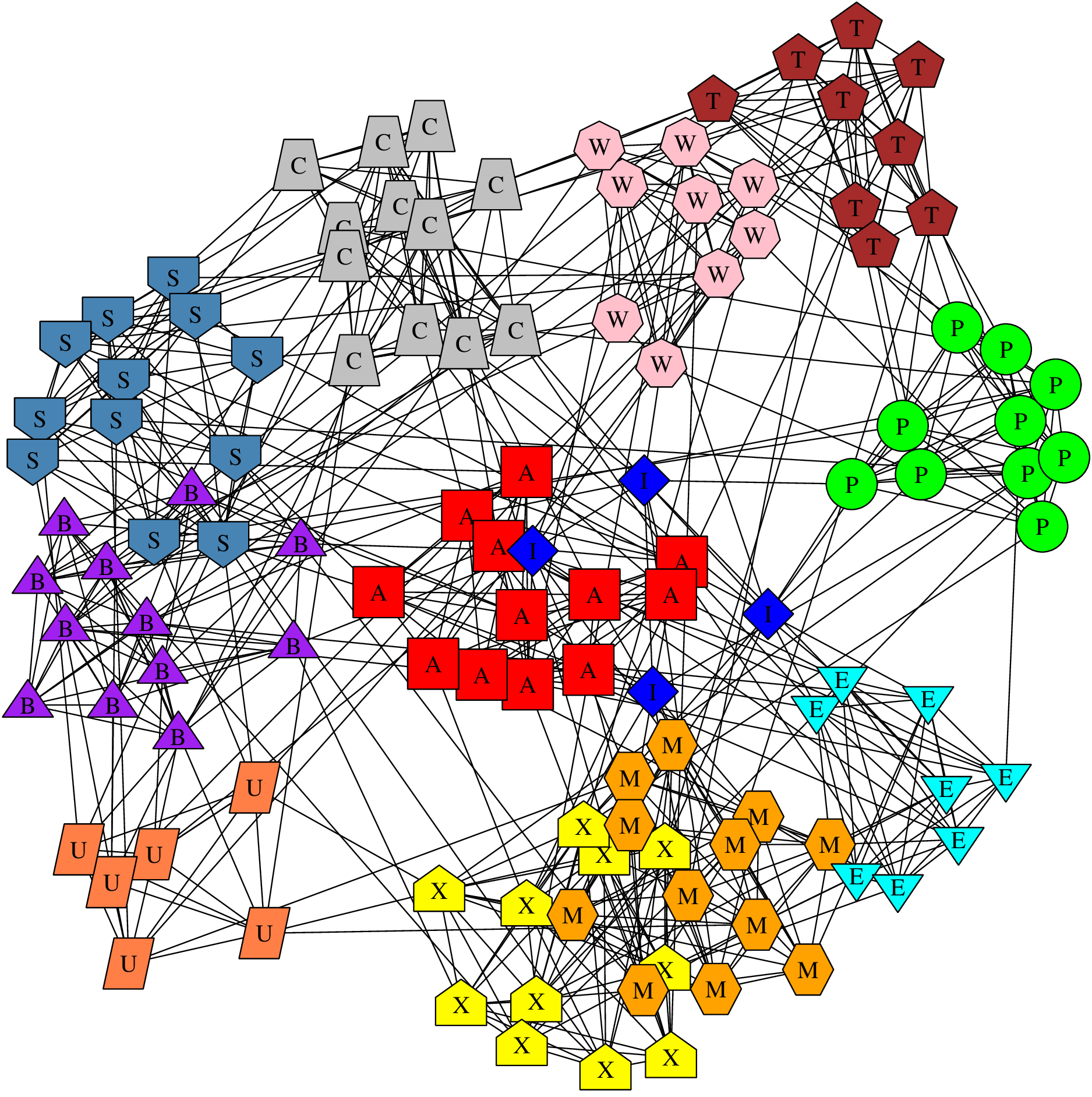}
	}
	\subfloat[]{
	    \label{NCAA_2005_3cycles.fig}
	    \includegraphics[width=0.48\textwidth]{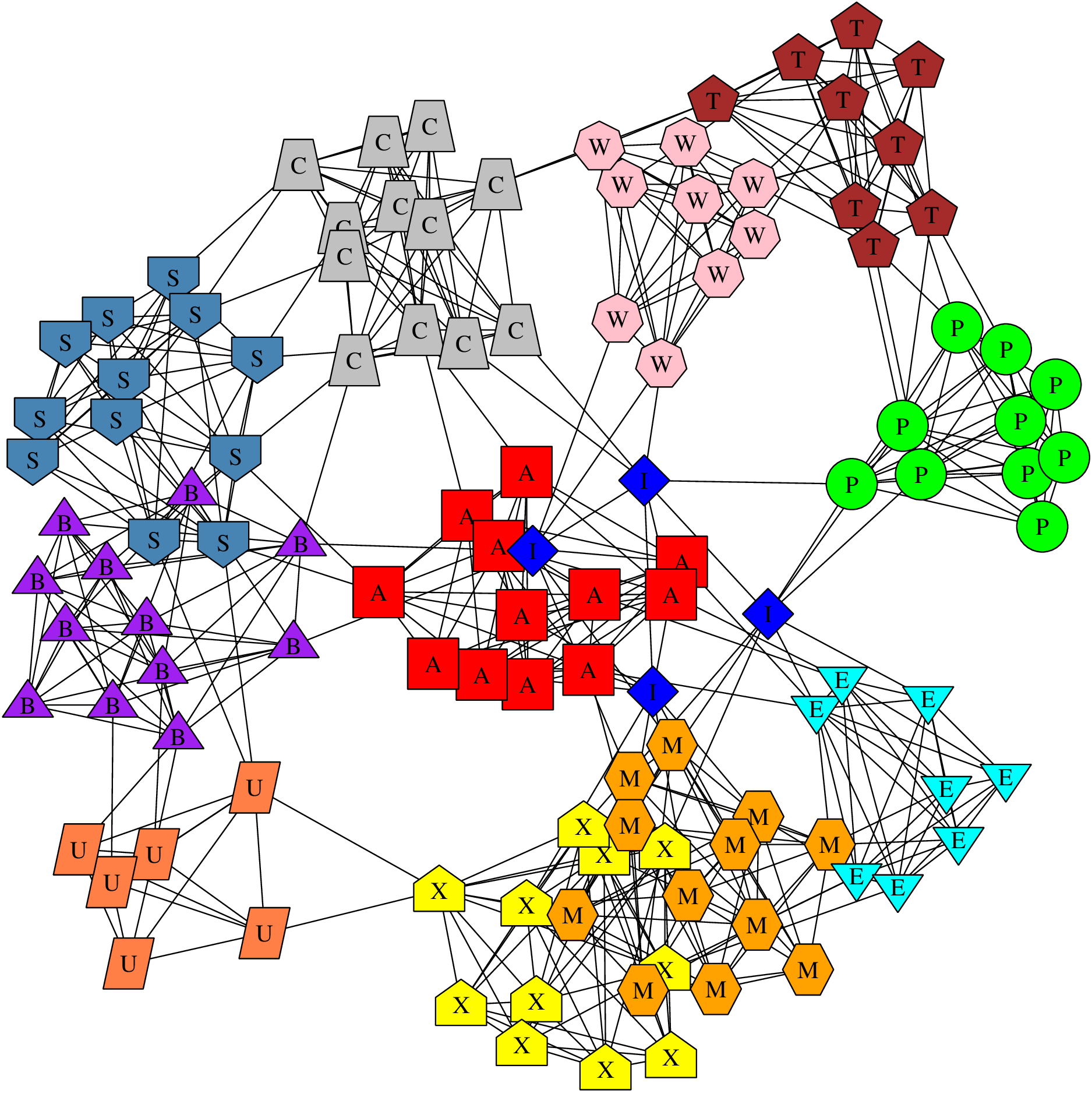}
	}
	\newline
	\subfloat[]{
	    \label{NCAA_2005_5not3.fig}
	    \includegraphics[width=0.48\textwidth]{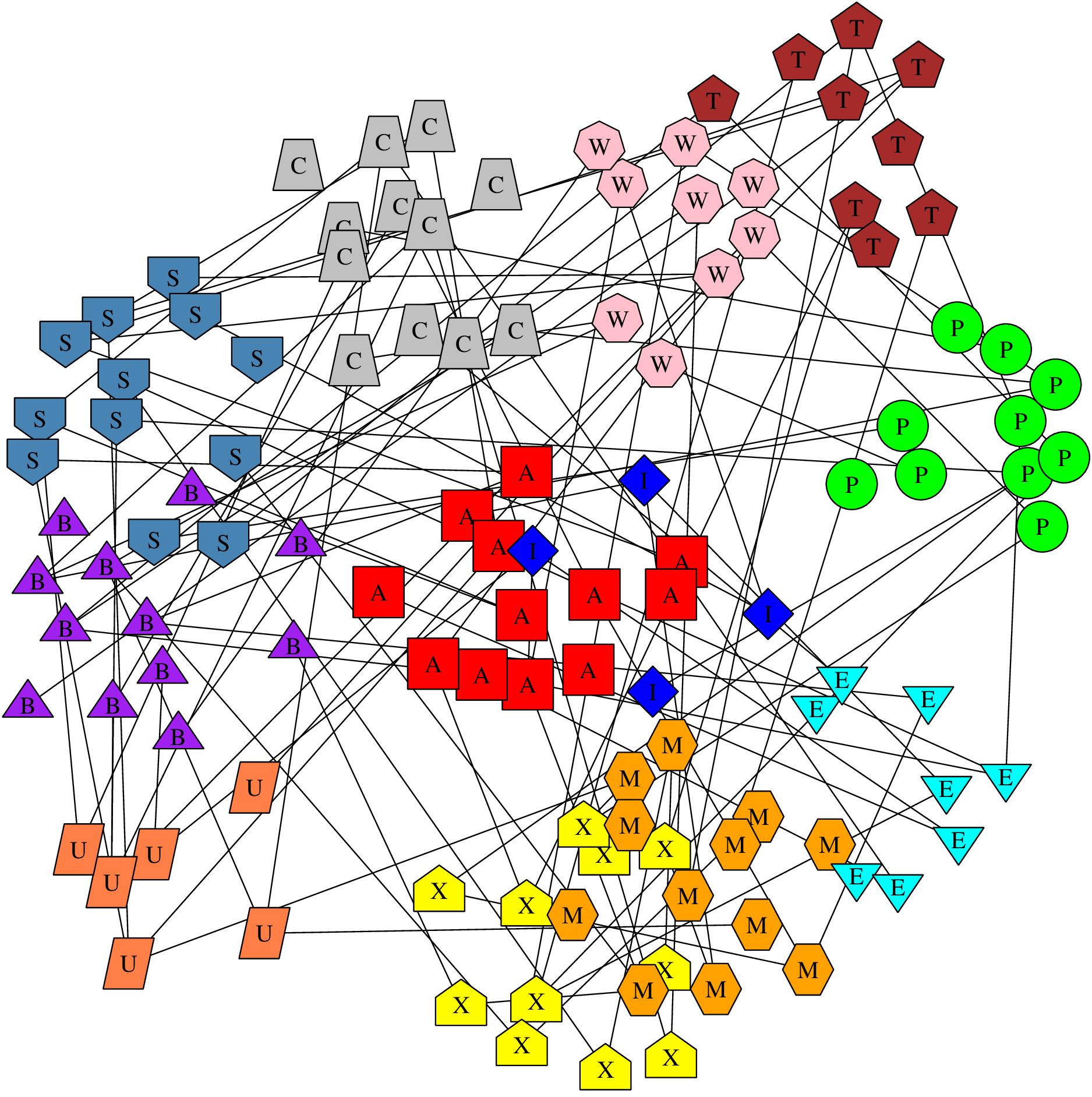}
	}
	\subfloat[]{
	    \label{NCAA_2005_5not3_newlayout.fig}
	    \includegraphics[width=0.48\textwidth]{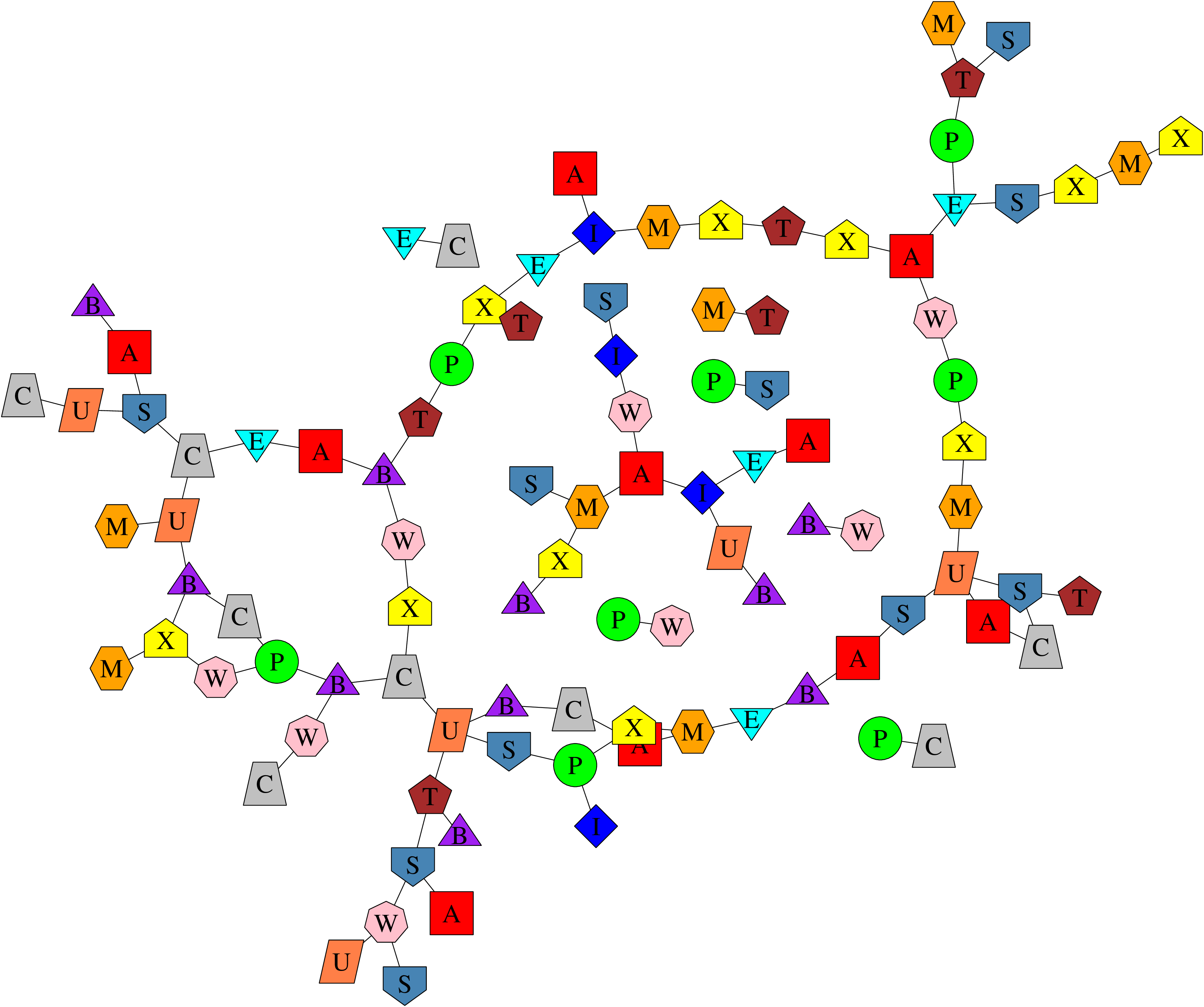}
	}
	\caption[]{(color online) The NCAA Div I-A 2005 regular season with all edges \protect{\subref{NCAA_2005_orig.fig}}, with 3-cycles only 
	\protect{\subref{NCAA_2005_3cycles.fig}}, and with just $C_{5 \setminus 3}$ edges \protect{\subref{NCAA_2005_5not3.fig}}.  
	\subref{NCAA_2005_5not3_newlayout.fig} is the same graph as \protect{\subref{NCAA_2005_5not3.fig}} but with a layout emphasizing that no edges 
	within conferences remain (degree zero nodes omitted). As per \protect{\cite{NewmanPark:Football}}, the conferences are: A = Atlantic Coast, B = Big 12, C = Conference USA, E = 
	Big East, I = Independent, M = Mid-American, P = Pacific Ten, S = Southeastern, T = Western Athletic, U = Sun Belt, W = Mountain 
	West, X = Big Ten. }
	\label{NCAA_2005.fig} % caption for the whole figure
\end{figure*}

Figure \ref{NCAA_2005_orig.fig} displays the original network, call it
$G$. As a first pass, let's use $j=\{3\}$ and generate $G_{3} =
\{V,C\}$, pictured in Figure \ref{NCAA_2005_3cycles.fig} using the
same layout as
\ref{NCAA_2005_orig.fig}.  This deletes all edges that do not participate 
in 3-cycles.  Most deleted edges are between conferences, though some
edges remain.  This will not split the network into seperate
components based on the communities but it
may be useful as a preprocessing step for betweenness or another
community detection algorithm.

In addition, let us build $C_{t \setminus s}$, as per Equation
\ref{eqn_t_not_s}.  For this network, we have chosen $t=5, s=3$.  
Figure \ref{NCAA_2005_5not3.fig} shows $G_{5 \setminus 3} = 
\{V, C_{t \setminus s}\}$, again using the same layout as 
\ref{NCAA_2005_orig.fig}.  For improved clarity, Figure 
\ref{NCAA_2005_5not3_newlayout.fig} shows $G_{5 \setminus 3}$ with 
a layout emphasizing that all edges are between conferences.  
	
\begin{figure*}
	\centering
		\subfloat[]{
		    \label{hist_NCAA.fig}
	    	\includegraphics[width=0.48\textwidth]{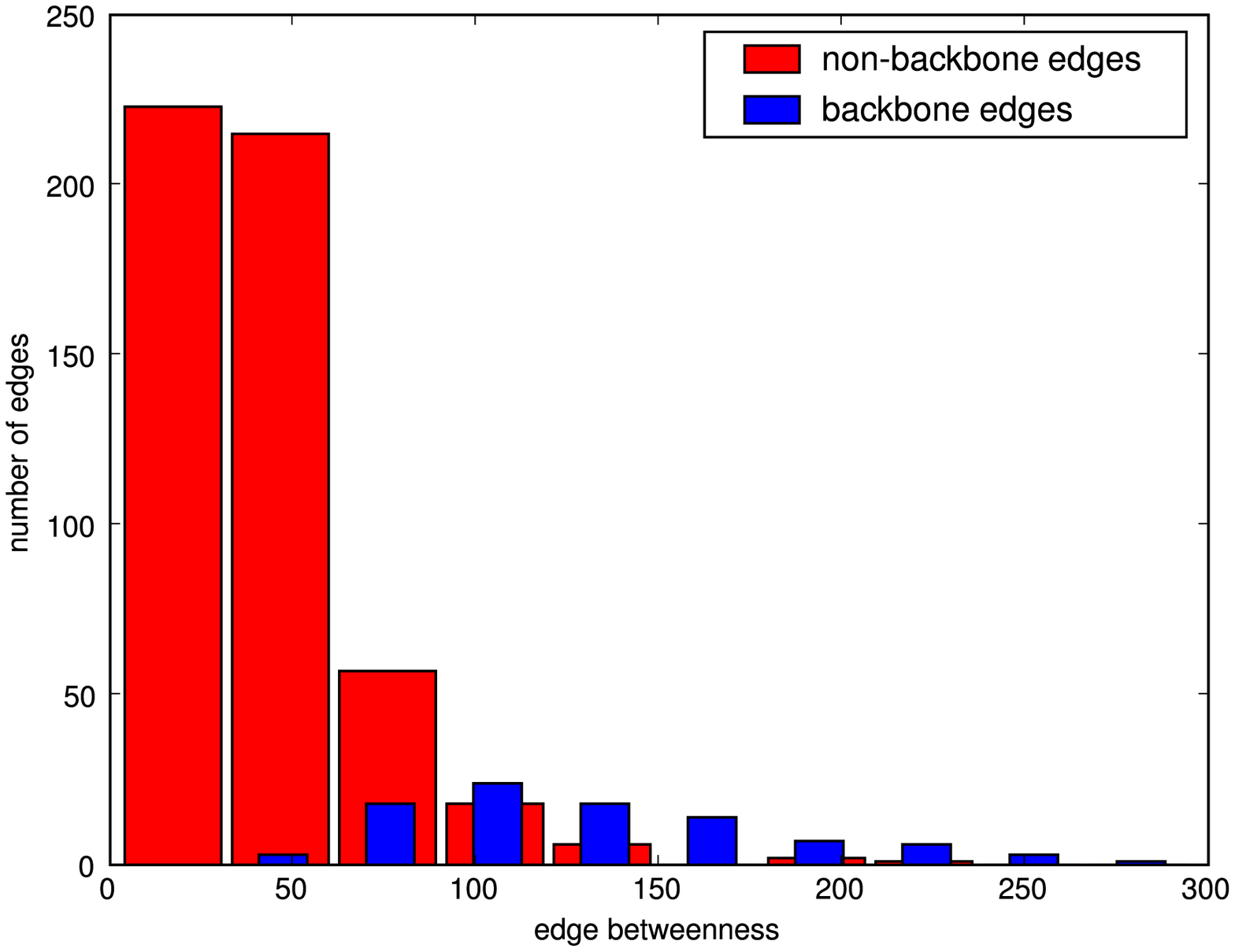}
		}
			\subfloat[]{
		    \label{hist_Sznajd.fig}
	    \includegraphics[width=0.48\textwidth]{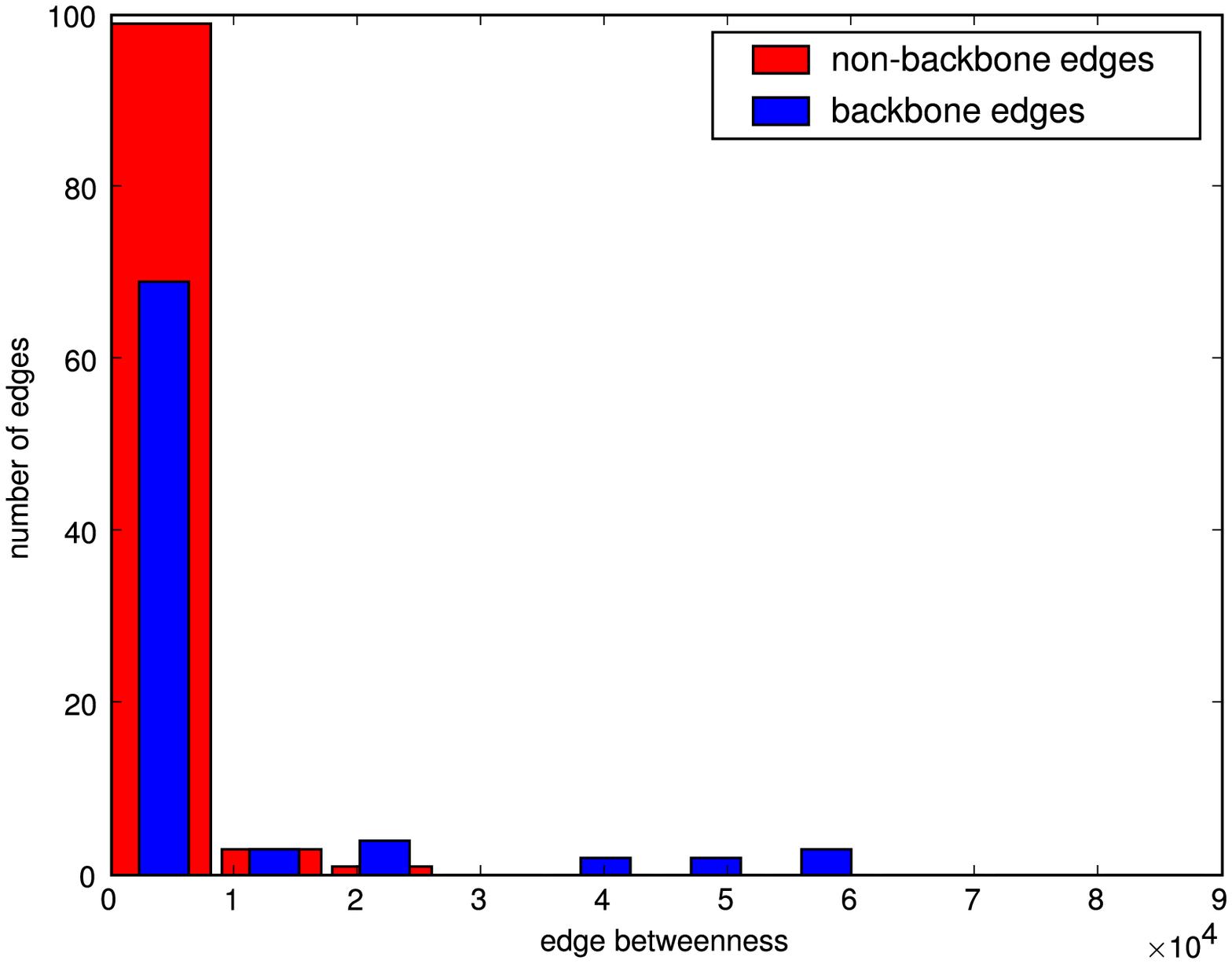}
		}
	\caption[]{(color online) Histogram of edge betweenness for
	non-backbone edges (red) and backbone edges (blue) for the
	NCAA 2005 football network \protect{\subref{hist_NCAA.fig}} and the
	Sznajd network shown in Figure~\protect{\ref{sznajd_net.fig}}
	\protect{\subref{hist_Sznajd.fig}}.  For the football network, the mean
	(unnormalized) betweenness is 42.8 for non-backbone edges and
	132.9 for backbone edges.  Note that backbone and non-backbone
	histograms use the same bins; the front-most bins have been
	narrowed for clarity.  The Sznajd non-backbone bins have also
	been scaled down by a factor of 25 for clarity. }
\end{figure*}

We propose that edges in $C_{5 \setminus 3}$ comprise the majority 
of this network's inter-community structure.  To test this, one can 
compare the distributions of edge betweenness for these backbone and 
non-backbone edges, as shown in Figure \ref{hist_NCAA.fig}.  Backbone 
edges tend to carry much higher betweenness values than the more 
common non-backbone edges.

\subsection{Sznajd Network}

One particularly interesting category of complex networks are the
so-called \emph{geographical models}
(e.g.~\cite{Gastner_Newman:2006,Costa_Diambra:2005}), whose nodes have
well-defined positions in an embedding metric space $S$.  Typically,
the connectivity in such networks is affected by the adjacency and/or
the distance between pairs of nodes, with nodes which are closer one
another having higher probability of being connected.  As an immediate
consequence of such an organizing principle, communities in
traditional geographical communites are closely related to the
presence of spatial clusters of nodes, i.e. groups of nodes which are
closer one another than with the rest of the network.  Introduced
recently, the family of geographical networks known as \emph{Sznajd
networks}~\cite{Costa_Sznajd:2005} allow rich community structure as a
consequence of running the Sznajd opinion formation
dynamics~\cite{Sznajd:2000} among the network edges instead of
considering the states associated to each network node.  Starting with
a traditional geographical network (called the \emph{underlying
network} $\Gamma$) where the connections are defined with probability
proportional to the distances between pairs of nodes, a percentage of
edges of $\Gamma$ are removed, yielding the initial condition for the
Sznajd dynamics.  Then, edges from $\Gamma$ are chosen randomly and
used to influence the respective surrounding connectivity.  For
instance, in case the chosen edge $(i,j)$ is on (i.e. it does
correspond to a link in the current growth stage), the edges in
$\Gamma$ which are connected to the nodes $i$ and $j$ are established
with probability $p$.  An analogue procedure is considered with
respect to edges which are absent.  In order to avoid convergence to
the trivial ground states where all edges are set on or off, the
dynamics also consider as feedback the total number of established
edges.

\begin{figure*}
	\begin{center}
	\subfloat[]{
		\includegraphics[width=0.48\textwidth]{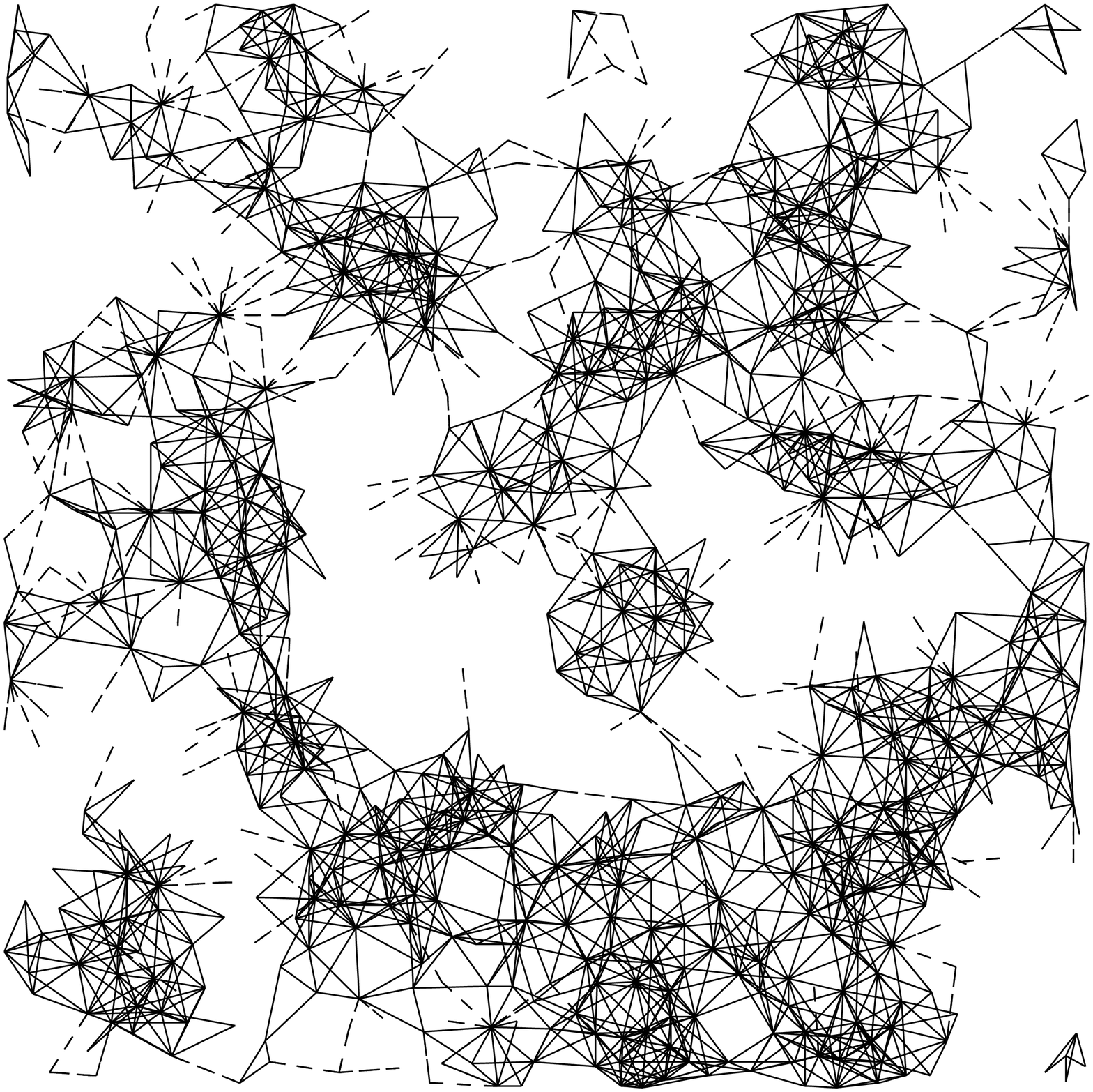} 
		\label{sznajd_3dashed.fig} 
	}
	\subfloat[]{
	    \label{sznajd_5not3bold.fig}
	    \includegraphics[width=0.48\textwidth]{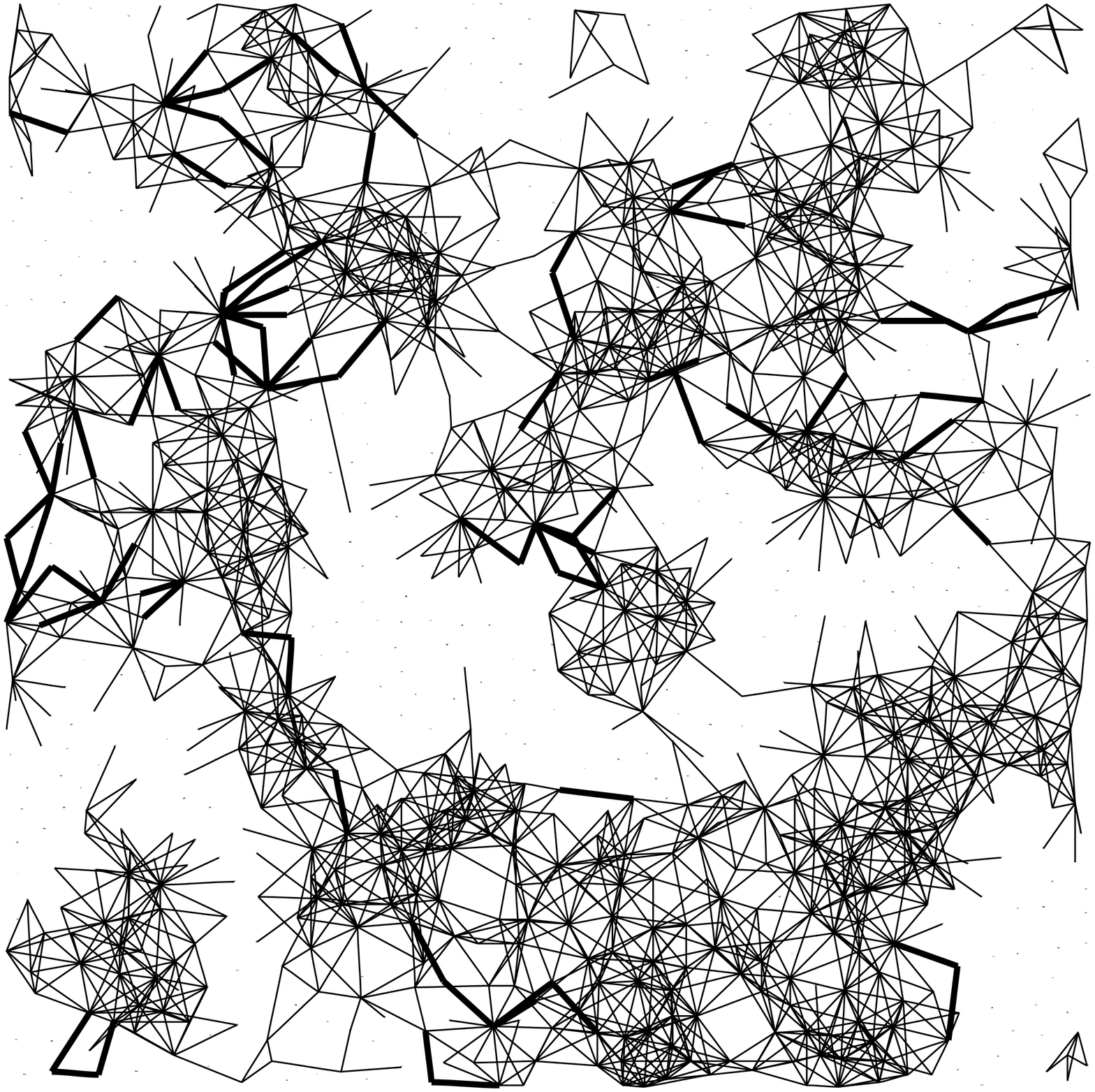}
	}
	\caption[]{A Sznajd network.  Edges that do not participate in 3-cycles are dashed \protect{\subref{sznajd_3dashed.fig}}. Edges in $C_{5 \setminus 3}$ are bold \protect{\subref{sznajd_5not3bold.fig}}.  Note that nodes of degree zero have been omitted for clarity. \label{sznajd_net.fig} } 
	\end{center}
\end{figure*}

Figure \ref{sznajd_3dashed.fig} shows a Sznajd Network.  Edges that do
not participate in 3-cycles are indicated.  As can be seen, many of
these edges fall ``outside'' of the more dense regions of the network.
This is a good first pass, and may be used to initialize another
algorithm, similar to our football result, but it will not give detailed
information on the hierarchical community structure.

Figure \ref{sznajd_5not3bold.fig} shows the same network as 
\ref{sznajd_3dashed.fig}, but with the edges of $C_{5 \setminus 3}$ 
highlighted.  One can imagine removing both the $C_3$ and 
$C_{5 \setminus 3}$ edges to further enhance the separation.
	
\section{Concluding Remarks}

The identification and characterization of the communities present in
complex networks stands out as one of the most important approaches
for understanding their structure and possible formation and
evolution.  At the same time, the distribution of cycles of various
lengths in a complex network has important implications for the
connectivity, resilience and dynamics of the respectively studied
networks.  The current work brought together these two important
trends, in the sense of applying short cycle detection as the means to
help the identification of communities in complex networks.  The
suggested methodology has been applied with promising results to the
identification of communities in a theoretical network model, more
specifically a Sznajd geometrical networks, as well as to a real-world
network (NCAA).  

The relationship between the cycles and communities in the
football network has been further investigated in terms of the
betweeness centrality measurement, confirming that the obtained
backbone edges tend to exhibit higher betweeness values.

\vspace{1cm}
{\bf Acknowledgments:}  L. da F. Costa thanks FAPESP (05/00587-5)
and CNPq (308231/03-1) for financial support.

\bibliographystyle{apsrev}

\end{document}